\documentstyle[aps,12pt,amssymb]{revtex}
\newcommand{\bm}[1]{\mbox{\boldmath $#1$}}

\def\DR{\rm I\kern-1.45pt\rm R}
\def\DC{\kern2pt {\hbox{\sqi I}}\kern-4.2pt\rm C}
\def\one{{\sf 1}\mkern-5.0mu{\rm I}}
\title{Asymptotic Ground State for 10 Dimensional Reduced Supersymmetric SU(2) Yang Mills Theory}
\author{G.M. Graf, J. Hoppe}
\address{Institute for Theoretical Physics, ETH H\"onggerberg, CH-8093 Z\"urich}
\begin{document}

\maketitle
\vspace{0.4cm}
\begin{abstract}
We calculate the power law decay, and asymptotic form of a (unique) SO(9) and SU(2) invariant wave function satisfying, to leading and sub-leading order, $Q_{\hat{\beta}} \psi = 0$ for all 16 supercharges of the matrix model corresponding to supermembranes in 11 space-time dimensions. 
\end{abstract}
\vspace{0.4cm}
Recently, a lot of attention has been devoted to the question, whether certain supersymmetric matrix models possess normalizable zero-energy states, or not \cite{c1}. These models are supersymmetric extensions of bosonic membrane matrix models \cite{c2} and were first studied in \cite{c3} (as reduced Susy Yang Mills theories) and \cite{c4} (as super-membrane matrix models), while broader interest in them exploded $1\frac{1}{2}$ years ago \cite{c5}. 

In this paper we will determine the asymptotic form (to leading, and sub-leading order) of a SO(9) and SU(2) invariant zero energy state and, in particular, calculate the power law decay at $\infty$ from the equations $Q_{\hat{\beta}} \psi = 0$, $\hat{\beta} = 1, \dots ,16$. 
Our results are in complete agreement with those of Halpern and Schwartz (cp. \cite{c1}) who, by a formidable 2$^{\mathrm{nd}}$ order calculation (i.e. using the Hamiltonian, instead of the supercharges) obtained the asymptotic power law decay, and form of the wave function. While it is reassuring that two different calculations give the same answer, the main reason for presenting our results is that the first order calculation is {\em much} simpler 
(the sub-leading part of the wave function follows from naive first order perturbation theory, no explicit canonical transformations decoupling bosonic and fermionic degrees of freedom are necessary, \dots; our method generalizes to SU(N), and gives closed expressions for all higher order corrections). 
We should mention, that we too started our investigations using $H$ (instead of $Q$), sometime last September (and refer to the Diploma thesis of Hasler, cp. \cite{c1}, for a Hamiltonian Born-Oppenheimer analysis of the $d=2$ SU(2)-model, as well as to \cite{c6}), while only recently we realized that, and how, the exact power law decay can be determined easily from the first order equations.

The ground state wave function is obtained from simple perturbation theory, using the supercharges $Q_{\hat{\beta}}=Q_{\hat{\beta}}^{(0)}+Q_{\hat{\beta}}^{(1)}+\dots $. Demanding SO(9) invariance, the $Q_{\hat{\beta}}^{(0)}$ have a unique (common) zero-mode, $\psi_0$, up to multiplication by a function of the SO(9) invariant variable $r$, that measures the distance from the origin in the space of configurations having vanishing potential energy. 
The full ground state is of the form (just as in $Q_{\hat{\beta}}$, the expansion parameter is $r^{-3/2}$) $\psi = f(r) \, (\psi_0 + \psi_1 + \dots )$, where $f(r)$ and $\psi_1$ can be determined from $Q_{\hat{\beta}}^{(1)} (f \psi_0) + Q_{\hat{\beta}}^{(0)} (f \psi_1 ) =0$, the sub-leading term in $Q \psi =0$.
\section{The model}
If the gauge group is SU(2), the model may be viewed as a supersymmetric version of a quantum mechanical system of 9 particles in  $\DR^3$ whose potential energy is the sum of the squares of twice the individual areas spanned by each pair of particles with the origin:
\begin{equation} H = - \sum_{s=1}^9 \, \vec{\nabla}_s^2 + \sum_{s<t} \left( \vec{q}_s \times  \vec{q}_t \right)^2 + i \vec{q}_s \cdot \vec{\Theta}_{\hat{\alpha}} \times \vec{\Theta}_{\hat{\beta}} \, \gamma^s_{\hat{\alpha}\hat{\beta}} \label{e1} \, . \end{equation}

The anti-commutation relations of the 48 (hermitian) fermionic operators $\Theta_{\hat{\alpha}A}$, $\hat{\alpha}=1,\dots,16$, 
$A=1,2,3$, 
are taken to be 
\begin{equation} \left\{  \Theta_{\hat{\alpha}A} ,\Theta_{\hat{\beta }B} \right\} = \delta_{\hat{\alpha} \hat{\beta}} \, 
\delta_{AB}   \, , \label{e2} \end{equation}
and the SO(9) $\gamma$-matrices , $\{ \gamma^s , \gamma^t  \} = 2 \delta^{st} \one $, will be chosen to be 
\begin{equation} \gamma^9 =\left( \begin{array}{cc}  \one & \ 0 \\ 0 
& - \one \end{array} \right) \, , \,  \gamma^8 = \left( \begin{array}{cc}  \ 0 &\ \one  \\ \ \one & \ 0\end{array} \right) \, , \quad  
\gamma^j =\left( \begin{array}{cc}  0 & i \Gamma^j\\  -i \Gamma^j& 0 \end{array} \right) \, \label{e3} \end{equation}
with $\Gamma^j$ purely imaginary, antisymmetric, $\{ \Gamma^j, \Gamma^k \} = 2 \delta_{jk} \one_{8 \times 8} $.
\section{Symmetries}
Up to operators that vanish on SU(2) invariant wave functions , $H$ is (twice) the square of any of the 16 hermitian operators (generating 
supersymmetry transformations) 
\begin{equation} Q_{\hat{\beta}}= \vec{\Theta}_{\hat {\alpha}} \, \left( -i \gamma^t_{\hat{\alpha}\hat{\beta}} \vec{\nabla}_t + \frac{1}{2} \,  
\vec{q}_s \times \vec{q}_t \gamma^{st}_{\hat{\beta}\hat{\alpha}} \right) \, . \label{e4} \end{equation}
In the representation (\ref{e3}) the $\gamma^{st} := \frac{1}{2} ( \gamma^s \gamma^t - \gamma^t \gamma^s )$ read
\begin{eqnarray} \gamma^{jk} = \left( \begin{array}{cc} \Gamma^{jk} & 0 \\ 0 & \Gamma^{jk} \end{array} \right) \,  ,\quad 
\gamma^{j8} = \left( \begin{array}{cc} i\Gamma^{j} & 0 \\ 0 & -i\Gamma^{j} \end{array} \right) \,  , \nonumber \\
\gamma^{j9} = \left( \begin{array}{cc} 0 & -i \Gamma^{j}  \\  -i \Gamma^{j} & 0 \end{array} \right) \,  ,\quad  \gamma^{jk} = \left( 
\begin{array}{cc} 0 & -\one \\ \one &0  \end{array} \right) \,  .  \label{e5} \end{eqnarray}
$Q_{\hat{\beta}} $ and $H$ commute with the 3 components of 
\begin{equation} \vec{J} = - i \, \left( \underbrace{\vec{q}_s \times \vec{\nabla}_s}_{\bf L} +\underbrace{ \frac{1}{2} 
\vec{\Theta}_{\hat{\alpha}} \times \vec{\Theta}_{\hat{\beta}} }_{\bf M} \right) \label{e6} \end{equation}
(generating SU(2) gauge transformations), while they transform as a (real) spinor and scalar, resp., under SO(9) transformations 
generated by 
\begin{equation} J_{st} = -i \, \left( \underbrace{\vec{q}_s \vec{\nabla}_t - \vec{q}_t \vec{\nabla}_s }_{{\bf L}_{st}} + \underbrace{\frac{1}{4} 
\vec{\Theta}_{\hat{\alpha}} \gamma^{st}_{\hat{\alpha}\hat{\beta}} \vec{\Theta}_{\hat{\beta}}}_{{\bf M}_{st}} \right) \, . \label{e7} 
\end{equation}
\section{Coordinates}
To study the asymptotic behavior of wave functions that could be annihilated by $H$ (hence, all the 
$Q$'s), it is useful to write 
\begin{equation} \vec{q}_s = r \vec{e} E_s + \vec{y}_s \label{e8} \end{equation}
with $\vec{y}_s\cdot \vec{e} = 0$,  $\sum_s \vec{y}_s E_s =0$, $ \sum_s E_s^2 = 1 = \vec{e}^{\, 2}$. As 
\begin{equation} y_{sB} = (\delta_{AB} - e_Ae_B ) \, (\delta_{st} - E_s E_t ) \, q_{tA} \label{e9} \end{equation}
is effectively of order $r^{- \frac{1}{2}}$ (as $r \to \infty$) , due to the bosonic potential $V$ in (\ref{e1}) being $\sum_{s<t} (\vec{q}_s 
\times\vec{q}_t )^2 = r^2 \, \sum_t \vec{y}_t^2 + \sum_{s<t} (\vec{y}_s \times \vec{y}_t )^2$, the leading order expressions for $\vec{\nabla}_t 
e_A$, $\vec{\nabla}_t E_s$ and $\vec{\nabla}_t y_{sB}$ can easily be calculated. Differentiating $(q_{sA}q_{sA'} ) \, e_{A'} = r^2 e_A$, resp. 
$(q_{sA}q_{s'A} ) \,  E_{s'} = r^2 E_s$, one gets 
\begin{eqnarray} \nabla_{tA} e_B &=& \frac{E_t}{r} \, ( \delta_{AB} - e_A e_B ) + \frac{1}{r^2} y_{tB} e_A + O ( \frac{y^2}{r^3} ) \, , \nonumber \\
  \nabla_{tA} E_s &=& \frac{e_A}{r} \, ( \delta_{st} - E_sE_t) +\frac{1}{r^2} y_{sA} E_t + O ( \frac{y^2}{r^3} ) \, , \nonumber \\
 \nabla_{tA} r &=& e_A E_t \, , \label{e10} \end{eqnarray}
while (\ref{e9}) implies 
\begin{eqnarray} \nabla_{tA} \, y_{sB}& =& ( \delta_{AB} - e_A e_B ) \, ( \delta_{st} - E_sE_t)\nonumber \\  & &- \frac{e_B E_t}{r} y_{sA} - 
\frac{e_AE_{s}}{r} y_{tB} -\frac{1}{r^2} (e_A e_B \vec{y}_s \vec{y}_t + E_sE_t y_{s'A}y_{s'B}) + O ( \frac{y^3}{r^3} ) \, . \label{e12} \end{eqnarray}
Between the $y_{sB} $ there exist $11= 9+3-1$ independent linear relations, so that only 16 of the 27 variables are independent. To 
be absolutely explicit, we could parameterize the two unit vectors , $e$ and $E$ as
\begin{eqnarray} \vec{e} = \left( \begin{array}{c} \sin \theta \cos \phi  \\ \sin \theta \cos \phi  \\ \cos \theta \end{array}   \right) \, , \quad
E = \left( \begin{array}{c} \sin \epsilon_8 \sin \epsilon_7 \dots \sin \epsilon_2 \sin \epsilon_1 \\ \sin \epsilon_8 \sin \epsilon_7 \dots 
\sin \epsilon_2 \cos \epsilon_1 \\ \sin \epsilon_8 \sin \epsilon_7 \dots \cos \epsilon_2 \\ \vdots \\ \sin \epsilon_8 \cos \epsilon_7 \\
\cos \epsilon_8 \end{array} \right) \, , \label{e13} \end{eqnarray}  and write
\begin{equation} \vec{y}_s = \sum_{\alpha=1}^8 \left( u_{\alpha} E^{(\alpha)}_s \vec{e}_{\phi} +v_{\alpha} E^{(\alpha)}_s \vec{e}_{\theta}  \right) \label{e14} \end{equation}
with $ \vec{e}_{\theta} = \partial_{\theta} \vec{e}$, $\vec{e}_{\phi} = \frac{1}{\sin \theta }  \partial_{\phi} \vec{e}$, $E_s^{(\alpha )} = \prod_{j=\alpha +1}^8 (\sin \epsilon_j )^{-1} \cdot \partial_{\epsilon_{\alpha}} E_s $, making 
\[ (q_{sB}) \rightarrow ( r, \theta, \phi , \epsilon_1, \dots , \epsilon_8 , u_1, \dots , u_8, v_1 , \dots , v_8 ) \]
a proper (local) change of coordinates.
It is easy to see that the part of $\nabla_{tA}$ containing derivatives with respect to $r,\theta , \phi, \epsilon_1, \dots ,\epsilon_8$ is (in 
leading order) 
\begin{equation} e_AE_t \partial_r + \frac{e_A}{r} E_s L_{st} + \frac{E_t}{r} e_B L_{BA} \, .  \label{e15} \end{equation}
As we will see, derivatives  with respect  to the remaining variables will, in that order $(\frac{1}{r})$, act  as $- \frac{1}{4r}$ times the 
fermionic part of (\ref{e1}) , yielding 
\begin{equation} \frac{-i}{4r} \,  ( \delta_{AB} - e_A e_B ) \, ( \delta_{st} - E_sE_t)  \, \left( \vec{\Theta}_{\hat{\alpha}} \times 
\vec{\Theta}_{\hat{\alpha'}} \right)_B  \gamma^s_{\hat{\alpha}\hat{\alpha}'} \, , \label{e16} \end{equation}
to be added to (\ref{e15}), and then multiplied by $\gamma^t_{\hat{\beta}\hat{\beta}'} \Theta_{\hat{\beta}'A}$, to give zero (up to 
terms cancelled by the 'potential part' in (\ref{e4}) ), when acting on the wave function. 

\section{Selecting terms}
As explained in the introduction, the ground state should, asymptotically, be of the form
\begin{equation}  \psi = r^{-\kappa} e^{- \frac{r}{2} \sum_s \vec{y}_s^2} | F \rangle \label{e17} \end{equation}
where $ e^{- \frac{r}{2} \sum_s \vec{y}_s^2} ( | F \rangle = | F_1 \rangle + | F_0 \rangle)$ is a ground state, calculated    to first order in 
perturbation theory, of $ - \Delta_y + r^2 y^2 + H_F$, where $\Delta_y = \vec{\nabla}_{y_s}^{\bot} \cdot \vec{\nabla}_{y_t}^{\bot} (\delta_{st} - E_sE_t)$ and
\begin{eqnarray} H_F &=& i \, ( r E_s \vec{e} + \vec{y}_s ) \, \left( \vec{\Theta}_{\hat{\alpha}} \times \vec{\Theta}_{\hat{\beta}} \right)  
\gamma^s_{\hat{\alpha}\hat{\beta}} \nonumber \\ &=& H^{(0)}_F + H'_F \label{e18} \, . \end{eqnarray}
Provided $|F \rangle $ is chosen to be SU(2) and SO(9) invariant (hence $\psi$), (\ref{e15}) can be replaced by 
\[ e_A E_t \partial_r + \frac{e_A}{r} M_{ts}E_s + \frac{E_t}{r} M_{AB} e_B \, , \]
which, up to a term $- e_A E_t \sum_s \vec{y}^2_s$, yields
\begin{equation} - e_A E_t \frac{\kappa}{r} + \frac{e_A}{r} \, (\vec{\Theta} \gamma^{ts} \vec{\Theta} ) E_s + 
\frac{E_t}{2r} \, \left( \vec{\Theta}_{\hat{\alpha}A } ( \vec{\Theta}_{\hat{\alpha}} \cdot \vec{e} ) -( \vec{\Theta}_{\hat{\alpha}} \cdot \vec{ 
e} )  \vec{\Theta}_{\hat{\alpha} A}  \right) \, . \label{e15p} \end{equation}
Contracted with $\gamma^t_{ \hat{\beta} \hat{\beta}'} \Theta_{\hat{\beta}'A}$ (cp. (\ref{e4})), as well as $\frac{1}{8} ( 
\vec{\Theta}_{\hat{\rho}} \cdot \vec{e} ) (E_{t'}\gamma^{t'})_{\hat{\rho}\hat{\beta}}$ (to project onto terms relevant for the 
calculation of $\kappa$) one gets
\begin{equation} - \frac{\kappa}{r} + \frac{1}{2r} \, E_{t'}M_{t't}^{\parallel} M_{ts}E_s + \frac{1}{8r} ( \vec{\Theta}_{\hat{\beta}} \cdot 
\vec{e} ) \, ( \vec{\Theta}_{\hat{\beta}} \times \vec{e} ) \, \frac{1}{2} \,    \vec{\Theta}_{\hat{\alpha}}  \times \vec{\Theta}_{\hat{\alpha}} 
\label{e19} \end{equation}
 from (\ref{e15p}) (with $M_{st}^{\parallel} := \frac{1}{4}      ( \vec{\Theta}_{\hat{\alpha}} \cdot \vec{e} ) 
\gamma_{\hat{\alpha}\hat{\beta}}^{st}  ( \vec{\Theta}_{\hat{\beta}} \cdot \vec{e} )$), and (using  that $\gamma = \gamma^s E_s $ 
squares to $\one$ and letting $\vec{\Theta}^{\bot}$ denote the components of $\vec{\Theta}$ orthogonal to $\vec{e}$)
\begin{equation} - \frac{i}{16 r} \,   \left( \vec{\Theta}_{\hat{\rho}}  \cdot \vec{e} \right)  \left( \vec{\Theta}_{\hat{\alpha}}  \cdot 
\vec{e} \right)  \, \vec{e} \cdot  \left( \vec{\Theta}_{\hat{\beta}'}^{\bot } \times \vec{\Theta}_{\hat{\alpha}'}^{\bot } \right) \cdot \left( 
\gamma_{\hat{\rho }\hat{\beta}}      \gamma_{\hat{\beta }\hat{\beta}'}^t      \gamma_{\hat{\alpha }\hat{\alpha}'}^t - 
\delta_{\hat{\rho} \hat{\beta}}      \gamma_{\hat{\alpha }\hat{\alpha}'} \right) \label{e20}
\end{equation} 
from (\ref{e16}). 

Having no more bosonic derivatives, one can facilitate the determination of $\kappa$ by choosing $E_s = \delta_{s9}$ (i.e. 
$\bm{\epsilon} =\bm{0}$) in ( \ref{e19}) and (\ref{e20}); so $\kappa$ must be an eigenvalue of 
\begin{eqnarray} - \frac{1}{2} \sum_t M_{t9}^{\parallel} M_{t9} + \frac{1}{8} \left( \vec{\Theta}_{\hat{\beta}}  \cdot \vec{e} \right)  \left( 
\vec{\Theta}_{\hat{\alpha}}  \cdot \vec{e} \right)  \left( \vec{\Theta}_{\hat{\beta}}^{\bot } \cdot \vec{\Theta}_{\hat{\alpha}}^{\bot } 
\right) \nonumber \\
- \frac{i}{16}  \left( \vec{\Theta}_{\hat{\rho}}  \cdot \vec{e} \right)  \left( \vec{\Theta}_{\hat{\alpha}}  \cdot \vec{e} \right)  \, \vec{e} 
\cdot  \left( \vec{\Theta}_{\hat{\beta}'}^{\bot } \times \vec{\Theta}_{\hat{\alpha}'}^{\bot } \right) \, \gamma^9_{\hat{\rho} \hat{\beta}} 
\, \sum_{\epsilon=1}^8  \gamma^\epsilon_{\hat{\beta} \hat{\beta}'} \gamma^\epsilon_{\hat{\alpha} \hat{\alpha}'}  \label{e21} 
\end{eqnarray}
for $|F_0 \rangle$, a SU(2) and SO(9) invariant state of lowest energy of $H_F^{(0)} $ - which we will now determine.

\section{Diagonalization of $H_F^{(0)}$}
Writing 
\begin{equation} H_F^{(0)} = r M_{\hat{\alpha}A, \hat{\beta} B } \Theta_ {\hat{\alpha}A} \Theta_{ \hat{\beta} B }  \label{e22} 
\end{equation}  one sees that the antisymmetric, purely  imaginary 48$\times$48 matrix
\begin{eqnarray} M_{\hat{\alpha}A, \hat{\beta} B } &=& (i \epsilon_{ABC} e_C ) ( \gamma^s_{\hat{\alpha} \hat{\beta}} E_s ) 
\label{e23} \\ &=& S_{AB} \gamma_{\hat{\alpha} \hat{\beta}}  \nonumber \end{eqnarray}
is the tensor-product of 2 matrices, the first one having eigenvalues $-1,+1,0$, the second one $+1$ (with multiplicity 8) and $-1$ (with 
multiplicity 8) , as $\gamma^2 = \one$, and $\mbox{Tr} \,  \gamma =0$. Apart from $\vec{e}$ (eigenvalue $0$), the normalized 
eigenvectors of $S$ are (up to a phase)
\begin{equation} \vec{n}_{\pm}  = e^{\mp i \delta } \,  \frac{1}{\sqrt{2}} ( \vec{e}_{\theta} \pm i \vec{e}_{\phi} ) \label{e24} 
\end{equation} (with eigenvalue $\mp 1$, $\vec{n}_+ \times \vec{n}_- = -i \vec{e} , \vec{e} \times \vec{n}_{\pm} = \mp i 
\vec{n}_{\pm} $) and in order to have $\vec{n}_{\pm}$ transform like a vector ($ \vec{e}_{\theta} , \,  \vec{e}_{\phi}$ do not), it will be 
convenient to choose $\delta$ such that $\vec{n}_+ \pm \vec{n}_-$ are eigenvectors of $q_{sA}q_{sB}$. The eigenvectors of 
$\gamma$  could be labeled $v^{(\pm , \alpha = 1 \dots 8)}$, $\gamma v^{\pm} = \pm v^{\pm} $, with 
\begin{equation} v_{\hat{\alpha} } ^{(+, \alpha )} (\bm{\epsilon} =\bm{0}) = \delta_{\alpha \hat{\alpha} } \,  , \quad v_{\hat{\alpha} } ^{(-, 
\alpha )} (\bm{\epsilon} =\bm{0}) = \delta_{\alpha+8, \hat{\alpha} } \, . \end{equation} 
Fortunately, it will be sufficient to restrict the discussion to $\bm{\epsilon}=0$, where 
\begin{equation}        H_F^{(0)} = i r \vec{e} \,  \left( \vec{\Theta}_{\alpha} \times \vec{\Theta}_{\alpha} - \vec{\Theta}_{\alpha+8} 
\times \vec{\Theta}_{\alpha+8}\right) \label{e26}  \, ,      \end{equation} and the transformation to annihilation and creation 
operators, suggested by $M w^* = - \nu w^*$ (when  $Mw = + \nu w$), diagonalizing $H_F^{(0)}$, may simply be taken to be 
\begin{eqnarray} \vec{\Theta}_{\alpha} &=& A_{\alpha} \vec{e} + \vec{n}_+ a_{1 \alpha} + \vec{n}_-  a_{1 \alpha}^{\dagger} 
\nonumber \\      \vec{\Theta}_{\alpha+8} &=& \tilde{A}_{\alpha} \vec{e} + \vec{n}_+ a_{2 \alpha}^{\dagger} + \vec{n}_- a_{2 \alpha}      
\label{e27} \end{eqnarray}
 with $A_{\alpha} = \frac{1}{\sqrt{2}} (   a_{0 \alpha} +  a_{0 \alpha} ^{\dagger} ) $, $\tilde{A}_{\alpha} =  \frac{i}{\sqrt{2}} (   a_{0 \alpha} -  a_{0 \alpha} ^{\dagger} ) $,
\begin{equation}      \left\{ a_{\mu \alpha} , a_{\nu \beta}^{\dagger} \right\} = \delta_{\mu \nu } \delta_{\alpha \beta} \, , \quad \alpha, 
\, \beta = 1 \dots 8 , \quad \mu , \, \nu = 0,1,2  \, ,  \label{e28}       \end{equation}
$\{ a,a\} = 0 =\{ a^{\dagger},a^{\dagger} \} $, 
\begin{equation} H_F^{(0)} = -16 r + 2r \sum_{{\mu' =1,2 } \atop \alpha}       \, a_{\mu ' \alpha} a_{\mu ' \alpha}^{\dagger} \, . 
\label{e29}         \end{equation}
For $\bm{\epsilon} \not= \bm{0} $, only the transformation (\ref{e27}) has to be changed ($a_{1\alpha} \to \vec{n}_- v_{\hat{\beta}}^{(+,\alpha )} \vec{\Theta}_{\hat{\beta}}, \, a_{2\alpha} \to \vec{n}_+ v_{\hat{\beta}}^{(-,\alpha )} \vec{\Theta}_{\hat{\beta}}$). - According to (\ref{e29}), 
\begin{equation}            | F_0 \rangle = \prod_{{\beta= 1 \dots 8} \atop {\nu' =1,2}} \,  a_{\nu' \beta}^{\dagger} | 0 \rangle_x \cdot | 
F_0^{\parallel} \rangle \,  , \label{e30}  \end{equation} (where  $a_{\nu' \beta} | 0 \rangle_x =0 \,  \forall \, \nu' ,  \alpha $ and $| 
F_0^{\parallel} \rangle $ is an arbitrary state formed out of the $a_0$'s) will have energy $-16r$, cancelling the zero-point energy 
coming from the 16 independent $y$-modes in $e^{- r y^2/2}$. As in $H_F' = i \vec{y} \cdot \vec{\Theta}_{\alpha}  \times 
\vec{\Theta}_{\beta}  \gamma_{\alpha \beta}^s $
 only those terms contribute, where in one of the two $\vec{\Theta}$'s the term 
$\parallel \vec{e} $ , i.e. $A_0$ (or $\tilde{A}_0$),
 is picked out, while in the other one $a_1$ or $a_2$ must be chosen, the relevant part of $H_F    $ will be linear in $a_1$ resp. $a_2$ (raising the energy by $2r$), as well as in one of the $y$-modes (again raising 
the energy by $2r$), so that all excited states reached via $H_F'$ will have the {\em same energy}, implying that 
\begin{equation}         | F_1 \rangle = - \frac{ 1}{4r} \, H_F' | F_0 \rangle  \, , \label{e31}     \end{equation}
which was anticipated in $(\ref{e16})$, where the $\frac{1}{r}$ contribution was written down that arises from taking derivatives with respect 
to the $y$-variables.
$|F_0^{\bot} \rangle = \prod_{{\beta = 1\dots 8} \atop {\nu'=1,2}} a_{\nu' \beta}^{\dagger} |0 \rangle_x$, first defined explicitly only at one point ($E_s = \delta_{s9}$ , resp. $\bm{\epsilon}= \bm{0}$), may be extended to a SO(9) invariant state (via $| F_0^{\bot} \rangle_{\bm{\epsilon}} = U(\bm{\epsilon}) | F_0^{\bot} \rangle_{{\footnotesize \bm{0}}} $) as $| F_0^{\bot} \rangle_{{\footnotesize \bm{0}}} $ is invariant under the little group SO(8) (leaving $\bm{\epsilon}=\bm{0}$ fixed) - SO(8) preserves the particle numbers $N_1 = \sum_{\alpha} a_{1 \alpha }^{\dagger} a_{1\alpha}$ and  $N_2 = \sum_{\alpha} a_{2 \alpha }^{\dagger} a_{2\alpha}$, and $| F_0^{\bot} \rangle_{{\footnotesize \bm{0} }} $ is the only state with $N_1 + N_2 = 16$. 
\section{SO(9) invariant $|F_0^{\parallel} \rangle $}
Denoting $a_{0 \alpha}^{\dagger}= \frac{1}{\sqrt{2}} \left( ( \vec{\Theta }_{\alpha} \cdot \vec{e} ) + i ( \vec{\Theta }_{\alpha +8} 
\cdot \vec{e} ) \right) $ by $\mu_{\alpha}$, $\alpha = 1 \dots 8$, $|F_0^{\parallel}\rangle $ will be of the form
\begin{equation}                | F_0^{\parallel} \rangle= ( p + p_{\alpha} \mu_{\alpha} + \frac{1}{2} p_{\alpha \beta} \mu_{\alpha} 
\mu_{ \beta} + \dots + p' \mu_1 \mu_2 \dots \mu_8 ) | 0 \rangle_{\parallel} \,  \label{e32}  \end{equation}
with coefficients $p_{\dots} $ that may (and will) depend on the bosonic variables. In terms of $\mu_{\alpha}$ and $\partial_{ 
\mu_{\alpha}}$ the generators $M_{st}^{\parallel} $ read
\begin{eqnarray} M_{98}^{\parallel} &=& \frac{i}{2}  \mu_{\alpha}  \partial_{ \mu_{\alpha}} - 2i \, , \quad     M_{9j}^{\parallel} =-i ( 
b_j + B_j) \nonumber \\ M_{ij}^{\parallel} &=& \frac{1}{2}  \mu_{\alpha} \Gamma^{ij}_{\alpha \beta} \partial_{ \mu_{\beta}}  \, , 
\quad     M_{8j}^{\parallel} =( B_j - b_j) \label{e33}    \end{eqnarray}
where 
\begin{equation}            b_j := \frac{i}{4} \mu_{\alpha} \Gamma^j_{\alpha \beta} \mu_{\beta}  \, , \quad B_j := - \frac{i}{4} 
\partial_{\mu_{\alpha}} \Gamma^j_{\alpha \beta} \partial_{\mu_{\beta}}  \label{e34} \end{equation}
change the $\mu$-degree by 2, (whereas SO(7) $\otimes $ U(1),  generated by the $M^{\parallel}_{ij}$ and $M^{\parallel}_{98}$, is 
realized linearly).

It is easy to see that, over $\mathbb{C}$, the 256 dimensional Hilbert space ${\cal H}$ of $\mu_{\alpha}$-polynomials does not contain 
any element annihilated by all 36 generators  (\ref{e33}). Rather , it splits into 3 SO(9) invariant subspaces (irreducible 
representations) of dimension 44, 84 and 128 ( all odd polynomials), and only the 44, for which
\begin{equation} |0 \rangle_{\parallel} \, , b_j |0 \rangle_{\parallel} \, ,   b_j b_k |0 \rangle_{\parallel} \, , B_j | \mu_1 \dots \mu_8 
\rangle , | \mu_1 \dots \mu_8 \rangle  \label{e35}                    \end{equation}
may serve as an explicit basis, occurs (as the symmetric traceless part in $9 \otimes 9$) in tensor products of the fundamental 
(vector) representation 9 of SO(9).
The only SO(9) invariant state is therefore
\begin{equation}     |  F_0^{\parallel} \rangle  = (E_s E_t - \frac{1}{9} \delta_{st} ) | 44; st \rangle \label{e36}          \end{equation}
which at $\bm{\epsilon}=\bm{0}$ is  
\begin{equation}          (1+ \frac{1}{4!} e_{\alpha\beta\gamma\epsilon} \mu_{\alpha} \mu_{\beta} \mu_{\gamma} \mu_{\epsilon} + 
\eta \mu_1 \mu_2 \dots \mu_8 ) | 0 \rangle_{\parallel} \, ,  \label{e37}      \end{equation}
the unique ($\eta=\pm 1$ only depends on which representation one chooses for
the $\Gamma^j$) SO(8) invariant state in ${\cal H}_{44}$ (${\cal H}_{44}$ decomposes into 3 irreducible representations of SO(8), $1 \oplus 
8_{v} \oplus 35_{v}$)
The quartic element ${\cal E}$ in (\ref{e37}) satisfies 
\begin{equation}          B_j {\cal E} = b_j \, , \quad b_j {\cal E} = \eta B_j \mu_1\mu_2 \dots  \mu_8      \, . \label{e38} \end{equation}
One has
\begin{equation}             [ B_j , b_k ] = \delta_{jk} \left( 1- \frac{1}{4} \mu_{\alpha} \partial_{\mu_{\alpha}} \right) + \frac{1}{4} (\Gamma^{jk})_{\alpha\beta} \mu_{\alpha}\partial_{\mu_{\beta}},  \, \label{e39}  
 \end{equation} $b_jb_k = \eta B_jB_k \mu_1\mu_2 \dots  \mu_8 $.

\section{Evaluation of $\kappa$}
Keeping in mind that (at $\bm{\epsilon} =\bm{0}$)
\begin{eqnarray} \vec{\Theta}_{\alpha}^{\bot} &=&      \vec{n}_+ a_{1 \alpha} + \vec{n}_-  a_{2 \alpha}^{\dagger} \nonumber \\ 
\vec{\Theta}_{\alpha + 8}^{\bot} &=&      \vec{n}_- a_{2 \alpha} + \vec{n}_+  a_{2 \alpha}^{\dagger} \, , \label{e27p} 
     \end{eqnarray}
it is straightforward to apply (\ref{e21}) to 
\begin{equation}         |F_0 \rangle = \prod a^{\dagger}_{\beta \nu'} | 0 \rangle_{\bot} \,  | F_0^{\parallel} \rangle \label{e23p}      
\end{equation}  (let us concentrate on the terms giving back $|F_0 \rangle $ ). The first contribution to $\kappa$ is
\begin{eqnarray} - \frac{1}{2} \sum_t M^{\parallel}_{t9}  M^{\parallel}_{t9}  &=& -  \frac{1}{2} \left(  \frac{1}{2}     \sum_{t,s=1}^9   
(M^{\parallel}_{ts})^2    -  \frac{1}{2}     \sum_{\alpha,\beta =1}^8   (M^{\parallel}_{\alpha\beta})^2   \right) \nonumber \\ &=&  - 
\frac{1}{2}  \left( C_9(44) - \hat{C}_8 \right) = +9  \label{e40} \, , \end{eqnarray} where we have used the fact that the SO(8) Casimir 
operator $\hat{C}_8 $ gives 0, when acting on (\ref{e37}), and\footnote{We are grateful to R.Suter for looking up the value of $C_9$, and for discussions concerning $|F_0^{\parallel} \rangle$.} $C_9(44)= -18$.
The second contribution to $\kappa $ is 
\begin{eqnarray} \frac{1}{8}  ( \vec{\Theta}_{\alpha} \cdot \vec{e} )  ( \vec{\Theta}_{\alpha} \cdot \vec{e} )   (   \vec{n}_- 
a_{1\alpha}  \vec{n}_+  a_{2 \alpha}^{\dagger} )      \nonumber  \\ + \frac{1}{8}  ( \vec{\Theta}_{\alpha+8} \cdot \vec{e} )  ( 
\vec{\Theta}_{\alpha+8} \cdot \vec{e} )   (   \vec{n}_+ a_{2\alpha}  \vec{n}_-  a_{1\alpha}^{\dagger} ) = 1 \label{e41}  
\end{eqnarray}
(when acting on $|F_0 \rangle$).
The third contribution to $\kappa$ is obtained by picking the terms
$ ( \vec{n}_- \times \vec{n}_+) a_{1 \alpha}^{\dagger} a_{1 \alpha}$ $(\hat{\alpha}' = \hat{\beta}' = \alpha ) $ and $ ( \vec{n}_+ \times \vec{n}_-) a_{2 \alpha}^{\dagger} a_{2 \alpha}$ $(\hat{\alpha}' = \hat{\beta}' = \alpha + 8) $ in $\vec{\Theta}_{\hat{\beta}'}^{\bot} \times \vec{\Theta}_{\hat{\alpha}'}^{\bot} $, i.e. (on $|F_0 \rangle$)
\begin{equation} \frac{1}{16} (\vec{\Theta}_{\hat{\rho}} \cdot \vec{e} )(\vec{\Theta}_{\hat{\alpha}} \cdot \vec{e} ) \, \gamma^9_{\hat{\rho} \hat{\beta}}\gamma^{\epsilon}_{\hat{\beta} \hat{\beta}'}  \gamma^9_{\hat{\beta}' \hat{\alpha}'}\gamma^{\epsilon}_{\hat{\alpha}' \hat{\alpha}} 
= -4 \, . \label{e42} \end{equation}
Altogether ,
\begin{equation} \kappa = 9 + 1 - 4 = +6 \label{e43} \end{equation}
making (\ref{e17}), i.e. $r^{-10} (r^4 e^{-r y^2/2} ) |F \rangle$, normalizable ($\int dr \, r^{10} (r^{-10})^2 < \infty$ ).
\section{Acknowledgement}
We would like to thank O.Augenstein, F.Finster, J.Fr\"ohlich, D.Hasler, J.Reinhardt,
R.Suter, and S.T.Yau for helpful discussions.

\end{document}